\documentclass[10pt,letterpaper]{article}
\pdfoutput=1
\usepackage{cogsci}
\usepackage{color}
\usepackage[hidelinks]{hyperref}
\usepackage{placeins}
\usepackage{natbib}
\usepackage{graphicx}
\usepackage{tikz}
\usepackage{amsmath, amssymb}
\usepackage{multirow}

\tikzset{
  treenode/.style = {shape=rectangle, rounded corners,
                     draw, align=center,
                     top color=white, bottom color=blue!20},
  root/.style     = {treenode, font=\Large, bottom color=red!30},
  env/.style      = {treenode, font=\ttfamily\normalsize},
  dummy/.style    = {circle,draw}
}

\newcommand{\gvec}[1]{\mbox{\boldmath$#1$}}
\newcommand{\gmat}[1]{\mbox{\boldmath$#1$}}

\title{Better safe than sorry:\\ Risky function exploitation through safe optimization}
 \author{\textbf{Eric Schulz$^1,$ Quentin J.M. Huys$^2,$ Dominik R. Bach$^3,$ Maarten Speekenbrink$^1$ \& Andreas Krause$^4$}\medskip\\ $^1$Department of Experimental Psychology, University College London, London, WC1H0AP\\
$^2$Translational Neuromodeling Unit, ETH and University of Z\"urich, Wilfriedstrasse 6, 8032 Z\"urich\\
$^3$Psychiatric Hospital, University of Z\"urich, Lenggstrasse 31, 8032 Z\"urich\\
$^4$Institute for Machine Learning, ETH Z\"urich, Universitaetstrasse 6, 8092 Z\"urich}

\begin{document}

\maketitle

\begin{abstract}
Exploration-exploitation of functions, that is learning and optimizing a mapping between inputs and expected outputs, is ubiquitous to many real world situations. These situations sometimes require us to avoid certain outcomes at all cost, for example because they are poisonous, harmful, or otherwise dangerous. We test participants' behavior in scenarios in which they have to find the optimum of a function while at the same time avoid outputs below a certain threshold. In two experiments, we find that Safe-Optimization, a Gaussian Process-based exploration-exploitation algorithm, describes participants' behavior well and that participants seem to care first about whether a point is safe and then try to pick the optimal point from all such safe points. This means that their trade-off between exploration and exploitation indicates intelligent, approximate, and homeostasis-driven behavior.\\

\end{abstract}
\stepcounter{footnote}\footnotetext{Corresponding author: eric.schulz@cs.ucl.ac.uk}

\section{Introduction}
Imagine you are hosting a dinner party. In the afternoon, you open up your fridge and kitchen cupboards to find a plethora of ingredients at your disposal. Aiming to amaze your friends with an unique culinary experience, you decide to prepare something extraordinary not found in recipe books. Considering your options, you generate expectations of how the tastes of different ingredients combine and interact to produce a -- hopefully memorable -- culinary experience. You have time to try out some options and experience their overall taste, learning about the effects of unusual combinations and methods of preparation. At the same time, however, you need to avoid certain combinations at all costs, for example those that are inedible, poisonous, or otherwise bad. 

This scenario is an example of a multi-armed bandit task \citep{srinivas2009gaussian}, where there are a number of actions or `arms' of the bandit (e.g., the possible dishes) which lead to initially unknown and stochastic outcomes or rewards (e.g., the taste of the dish), which are related to a set of features (e.g., the ingredients, the method of preparation, etc.). Through experience, one can learn the function which maps the features to the rewards and maximize the overall rewards gained over repeated plays of the bandit. A key issue in optimal behavior in such tasks is known as the exploration-exploitation dilemma: should I take an action which I know will lead to a high reward, or try an unknown action to experience its outcome and thereby learn more about the function mapping features to rewards, increasing my chances of gaining higher rewards in the future? In order to avoid certain bad outcomes (e.g., poisonous dishes), one should only explore uncertain options which are likely to be `safe'. Such restricted exploration-exploitation problems are ubiquitous in daily life, from choosing which restaurant to visit, which car to buy, all the way to whom to befriend. In our previous research on human behavior in contextual multi-armed bandits \citep{schulzexploration, schulzlearning}, we found that participants' behavior is well-described by Gaussian Process regression, a non-parametric regression tool that adapts its complexity to the data at hand by the means of Bayesian posterior computation. 

The aim of the present study is to assess how people behave when they have to maximize their rewards whilst avoiding outcomes below a given threshold. The task is couched as a function learning task, where participants choose an input and observe and accrue the output of the function. In two experiments with a uni- and bivariate function, we find that participants efficiently adapt their exploration-exploitation behavior to risk-inducing situations. Overall, they are well-described by a Gaussian Process-based safe optimization algorithm that tries to safely expand a set of `explorers' while simultaneously maximizing outputs within a set of possible `maximizers; \citep{sui15icml}. Such behavior might be based on the principle of homeostasis maintenance \citep{korn2015maintaining}, where organisms need to forage for food while avoiding the probability of starvation. Additionally, we find evidence that participants first assess whether points are safe and then attempt to maximize within this safe subset. This simplification of the task in terms of subgoals resonates well with recent results on approximate planning strategies in complex dynamic tasks \citep{huys2015interplay}. 

\section{Modeling learning and optimization}
If the task is to learn and maximize an unknown function, then two ingredients are needed: (a) a model to represent an unknown function, for which we will use Gaussian process regression, and (b) a method to safely choose the next inputs, for which we will use a safe optimization algorithm.

\subsection{Learning a function}
We assume people represent and learn a function through Gaussian process regression, a universal function learning algorithm which has been supported in previous research \citep{griffiths2009modeling,schulzlearning}.\\
A Gaussian Process ($\mathcal{GP}$) is a stochastic process of which the marginal distribution of any finite collection of observations is multivariate Gaussian \citep{rasmussen2006gaussian}. It is a non-parametric Bayesian approach towards regression problems and can be seen as a rational model of function learning as it adapts its complexity to the data encountered. Let $f(\gvec{x})$ be a function mapping an input $\gvec{x} = (x_1,\ldots,x_d)^\top$ to an output $y$. A $\mathcal{GP}$ defines a distribution $p(f)$ over such functions. A $\mathcal{GP}$ is parametrized by a mean function $m(\gvec{x})$ and a covariance (or kernel) function, $k(\gvec{x},\gvec{x}')$:
\begin{align}
m(\gvec{x})&=\mathbb{E}\left[f(\gvec{x})\right]\\
k(\gvec{x},\gvec{x}')&=\mathbb{E}\left[(f(\gvec{x})-m(\gvec{x}))(f(\gvec{x}')-m(\gvec{x}'))\right]
\end{align}
At time $t$, we have collected observations $\mathbf{y}_{1:t} = [y_1,y_2,\dots,y_t]^\top$ at inputs $\gvec{x}_{1:t}=(\gvec{x}_1,\dots,\gvec{x}_t)$. For each outcome $y_t$, we assume 
\begin{equation}
y_t=f(\gvec{x}_t)+\epsilon_t \quad \quad \epsilon_t \sim \mathcal{N}(0,\sigma^2)
\end{equation}
Given a $\mathcal{GP}$ prior on the functions
\begin{align}
f(\gvec{x}) \sim \mathcal{GP}\left(m(\gvec{x}),k(\gvec{x},\gvec{x}')\right),
\end{align}
the posterior over $f$ is also a $\mathcal{GP}$ with
\begin{align}
m_t(\gvec{x})&=k_{1:t}(\gvec{x})^\top(\gmat{K}_{1:t}+\sigma^2 \gmat{I}_t)\gvec{y}_{1:t}\\
k_t(\gvec{x},\gvec{x}')&=k(\gvec{x},\gvec{x}')-\gvec{k}_{1:t}(\gvec{x})^\top(\gmat{K}_{1:t}+\sigma^2 \gmat{I}_t)^{-1}\gvec{k}_{1:t}(\gvec{x}')
\end{align}
where $\gvec{k}_{1:t}(\gvec{x})=[k(\gvec{x}_1,\gvec{x}), \dots, k(\gvec{x}_t,\gvec{x})]^\top$, $\gmat{K}_{1:t}$ is the positive definite kernel matrix $[k(\gvec{x}_i,\gvec{x}_j)]_{i,j = 1,\ldots,t}$, and $\gmat{I}_t$ is a $t$ by $t$ identity matrix. This posterior distribution can be used to derive predictions for each possible input $\gvec{x}$ on the next time point, which again follow a Gaussian distribution. A key aspect of a $\mathcal{GP}$ is the covariance or kernel function $k$. The choice of a kernel function corresponds to assumptions about the kind of functions a learner expects. Here, we will use a squared exponential kernel:
\begin{align}
k_{\text{sqe}}(\gvec{x},\gvec{x}')&=\theta_1^2\exp\left(-\frac{(\gvec{x}-\gvec{x}')^2}{2\theta_2^2}\right)
\end{align}
This kernel induces a universal function learning engine and has been found to describe human function learning well \citep{griffiths2009modeling}. 
\subsection{Optimizing a function}
Given a learned representation of a function at time $t$, this knowledge needs to be used to choose a next input at time $t+1$. This is done through an acquisition function that takes the expected output for each input and the associated uncertainty to balance exploration and exploitation \citep{brochu2010tutorial}.\\
An algorithm that is well-poised to cope with the additional requirement to avoid outcomes below a threshold first separates possible inputs into those that are likely to provide outputs above the threshold (the safe set) and those that are not, and then separates this safe set further into a set of maximizers (inputs that are likely to provide the maximum output) and expanders (inputs that are likely to expand the safe set). Following \cite{berkenkamp2015safe}, we define upper and a lower bounds of a confidence interval as sum of the current expectation $m_{t-1}$ and its attached uncertainty $\sigma_{t-1}$.
\begin{align}
u_{t}(\gvec{x})&=m_{t-1}(\gvec{x})+\beta_t \sigma_{t-1}(\gvec{x})\\
l_t(\gvec{x})&=m_{t-1}(\gvec{x})-\beta_t \sigma_{t-1}(\gvec{x}).
\end{align}
the parameter $\beta_t$ determines the width of the confidence bound, and we set it to $\beta_t=3$ to assure high safety a priori (i.e. 99.9\%). Using these bounds, we can define the safe set as all the input points in the set $\mathcal{X}$ of available inputs that are likely to lead to output values above the safe threshold, $J_{\text{min}}$
\begin{align}
\mathcal{S}_t=\{\gvec{x} \in \mathcal{X}|l_t(\gvec{x}) \geq J_{\text{min}}\}
\end{align}
The set of potential maximizers contains all safe inputs that are likely to obtain the maximum output value; these are the safe inputs for which the upper confidence bound $u_t$ is above the best lower bound:
\begin{align}
\mathcal{M}_t=\{\gvec{x} \in \mathcal{S}_t | u_t(\gvec{x}) \geq \text{max}_{\gvec{x}' \in \mathcal{X}}l_t(\gvec{x}') \}
\end{align}
To find a set of expanders, we define
\begin{align}
g_t(\gvec{x})=|\{\gvec{x}' \in \mathcal{X} \setminus \mathcal{S}_t|l_{t,(\gvec{x},u_t(\gvec{x}))}(\gvec{x}')\geq J_{\text{min}}\}|
\end{align}
where $l_{t,(\gvec{x},u_t(\gvec{x}))}(\gvec{x}')$ is the lower bound of $\gvec{x}'$ based on past data and a predicted outcome for $\gvec{x}$ which provides a new upper bound $u_t(\gvec{x})$. The function is used to determine how many inputs are added to the safe set after choosing input $\gvec{x}$ and observing the output it provides. This function is positive only if the new data point has a non-negligible chance to expand the safe set. The set of possible expanders is then defined as 
\begin{align}
\mathcal{G}_t=\{ \gvec{x} \in \mathcal{S}_t| g_t(\gvec{x}) \geq 0 \}
\end{align}
Normally, the safe optimization routine picks as the next point a safe point that is within the intersection of expanding and maximizing points, but currently shows the highest uncertainty measured by the difference between the upper and the lower bound. However, for this first investigation of human behavior within safe exploration scenarios we will focus on how participants choices of input points are influenced by simple membership of the 3 sets and if their behavior can be described by more heuristic, stepwise decision behavior.

\section{Experiment 1: Univariate functions}

\begin{figure}[ht!]
\label{fig:screen1}
  \centering
    \includegraphics[scale=0.45]{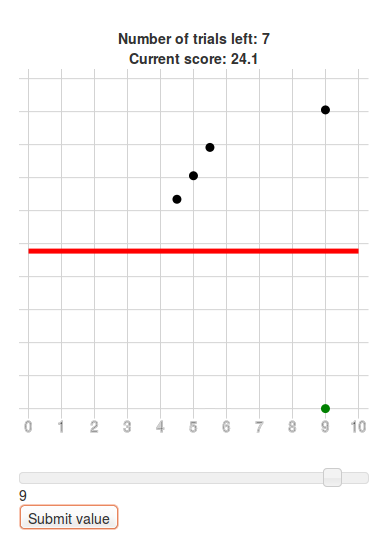}
\caption{Screenshot of first experiment.}
 \label{denseb}
\end{figure}

The first experiment required participants to maximize unknown univariate functions $f:x\rightarrow y$. On each trial $t = 1,\ldots,10$ in a block, they could choose an input value $x \in \{0,0.5,1,\dots,10 \}$ to observe (and accrue) an output  $y=f(x)+\epsilon$ with noise term $\epsilon \sim \mathcal{N}(0,1)$. The underlying functions were sampled from a $\mathcal{GP}$ with a squared exponential kernel ($l$=1, $\theta$=1). The objective was to maximize the sum of the obtained outputs over all trials in a block. A threshold $J_{\text{min}}$ was introduced and a block was ended abruptly if an output below this threshold was obtained. On average, that threshold was fixed to separate 50\% of the points into safe and unsafe points. Before the first trial, an initial safe point above the threshold was provided. A screenshot is shown in Figure~\ref{fig:screen1}.

\subsection{Participants}
61 participants (36 female) with an average age of 32.95 (SD = 8.02) were recruited via Amazon Mechanical Turk and received \$1 for their participation and a bonus of up to \$1, in proportion to their overall score.
\subsection{Procedure}
Participants were told that they had to maximize an unknown function while at the same time trying to avoid sampling below the red line as this would end the current block. After reading the instructions and performing an example task, they had to correctly answer 4 questions to check their understanding, then performed the task, and at the end saw their total score.

\subsection{Results}
\begin{figure}[ht!]
  \centering
    \includegraphics[scale=0.4]{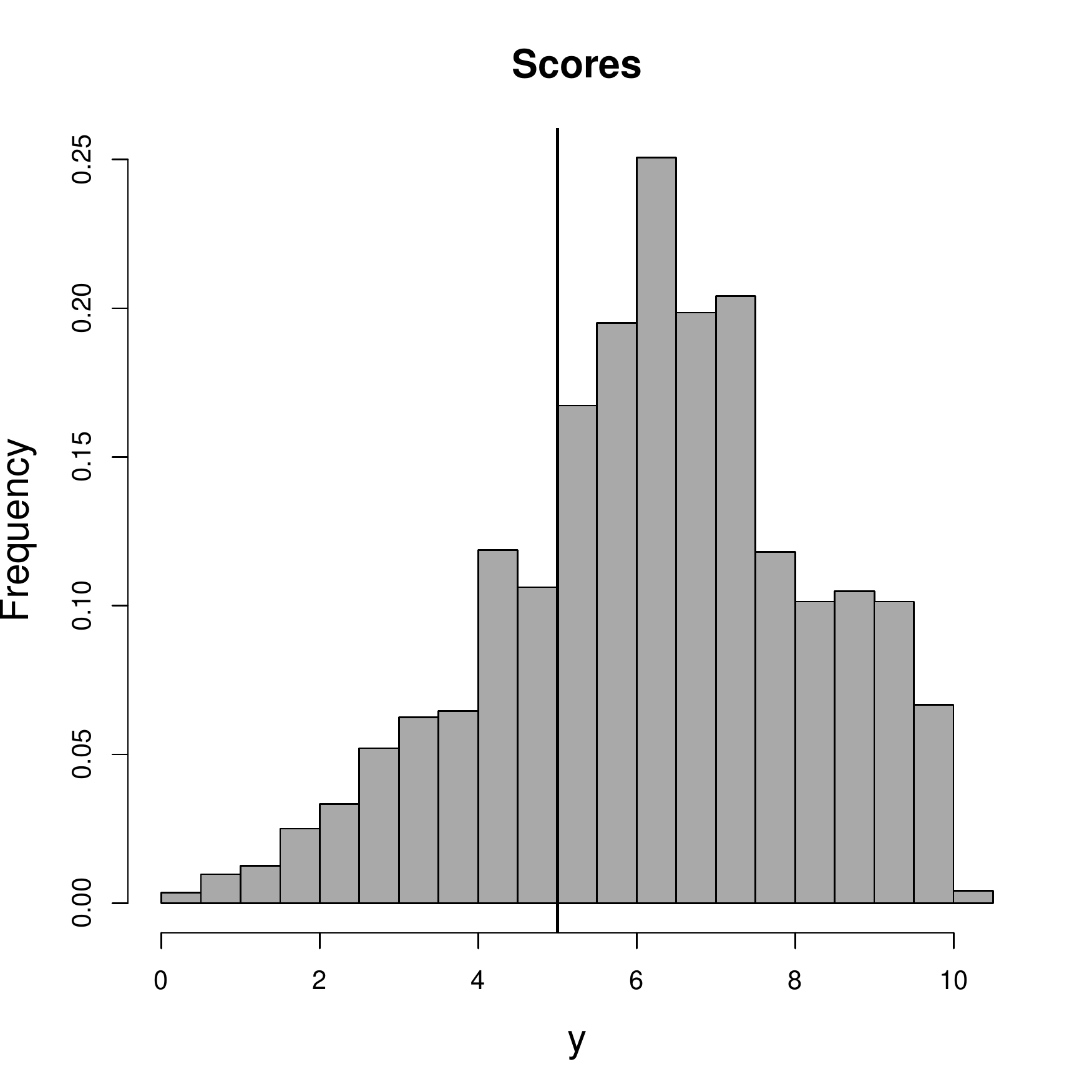}
\caption{Scores per trial. Black line: chance level.}
\label{fig:raw1}
\end{figure}

\begin{figure}[ht!]
\vspace{0.2cm}
\begin{tikzpicture}
  [
    grow                    = right,
    sibling distance        = 6em,
    level distance          = 9.5em,
    edge from parent/.style = {draw, -latex},
    every node/.style       = {font=\footnotesize},
    sloped
  ]
  \node [root] {root}
    child { node [env] {ignore}
      edge from parent node [below] {$p(sn)<0.99$} }
    child { node [dummy] {}
      child { node [env] {ignore}
              edge from parent node [above, align=center]
                {$p(mn)<0.05$}}
     child { node [env] {sample}
              edge from parent node [above, align=center]
                {$p(mn)>0.05$}}
              edge from parent node [above] {$p(sn)>0.99$} };
\end{tikzpicture}
\caption{Multi-level decision tree minimizing log-loss.}
\label{DT1}
\end{figure}
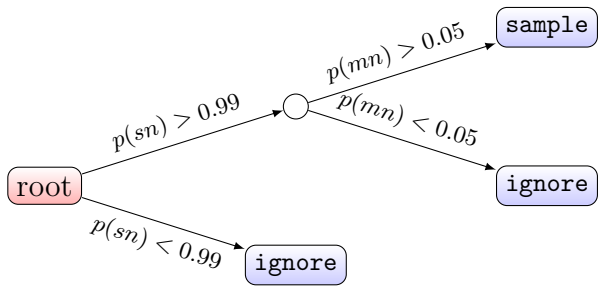

\begin{figure}[ht!]
  \centering
    \includegraphics[scale=0.4]{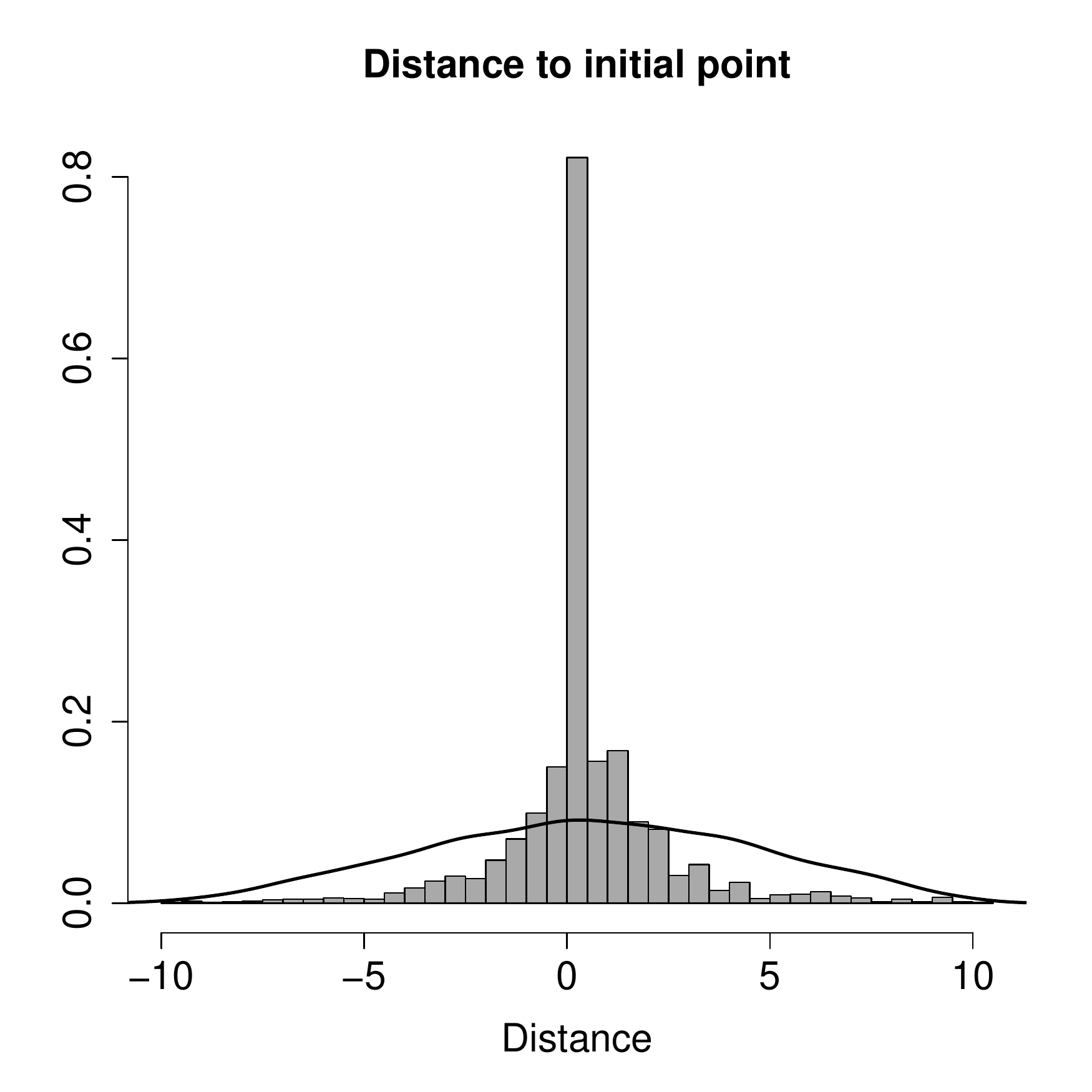}
\caption{Distance of chosen input points to initially provided point. Black line indicates expected density for sampling at random.}
\label{fig:dist1}
\end{figure}

As shown in Figure~\ref{fig:raw1}, participants obtained outputs higher than expected by chance on the large majority of trials and indeed the average score per participant was significantly higher than chance, $t(60)=13.311$, $p<0.01$.
In addition, the average number of trials per block statistically exceeded what would be expected if participants chose completely at random, $t(548) = 5.1201$, $p<0.01$ and participants' scores were positively correlated with trials ($r=0.25$, $p<0.01$). Taken together, these results indicate that participants learned the task and tended to chose safe inputs.\\
We used mixed-effects logistic regression analysis to asses which factors influenced participants' choices. The dependent variable was whether each input was chosen or not on each trial for each participant. As predictors, we used indicator variables for membership of an input of the safe, maximization, and expander set. Results indicated that the most plausible model was a model that contains all variables as fixed effects and a participant-specific random intercept, indicating that participants were influenced by set membership in an overall similar fashion. The coefficients of the fixed effects are presented in Table~\ref{results1} below.
\begin{table}[ht!]
\centering
\caption{Fixed effects estimate. Significant estimates are flagged.}
\vspace{0.2cm}
\label{results1}
\begin{tabular}{lll}
\hline
Variable& $b$ & $SE(b)$\\ \hline
Intercept&$-4.26^*$&$0.04$\\
Safe set&$1.57^*$&$0.06$\\
Maximizer set&$1.72^*$&$0.05$\\
Expander set&$0.12$&$0.05$\\
\hline
\end{tabular}
\end{table}
Comparing the magnitude of the slopes of the predictors, we can conclude that participants cared about all of the sets, but mostly about whether or not a point was safe and/or a maximizer.\\
Next, we used a random intercept decision tree analysis \citep{sela2011reemtree} to assess whether participants might utilize a simple but effective heuristic strategy that can be implemented as a decision tree, as suggested by \cite{huys2012bonsai}. For this, we replaced the indicators of set membership with probability assessments, substituting membership of the maximizer set with the probability of improvement, the safe set with the probability of being above the threshold, and the expander set with the probability of safely expanding the set (assessed through one step ahead forward simulation). The probability of improvement is defined as the probability that an input $\gvec{x}$ produces a higher output than the input $\gvec{x}^+$ that is currently thought to provide the maximum, and can be calculated as
\begin{align}
PI_t(\gvec{x})&=P\left(f(\gvec{x}) \geq f(\gvec{x}^+)\right)\\
&=\Phi\left(\frac{m_t(\gvec{x})-f(\gvec{x}^+)}{\sigma_t(\gvec{x})}\right)
\end{align}
where $\Phi$ is the cumulative Normal distribution function. Figure~\ref{DT1} depicts the decision tree which best fitted participants' choices. This analysis shows that participants seem to partition the problem into two sub-goals: first they conservatively assess whether or not a point is safe, then they maximize within that safe set. That expanders are not considered within this decision process could be due to the brevity of the task (10 trials) or to risk aversion (i.e., the fear of sampling below the threshold).
The non-inclusion of possibly expanding points also means that participant only tried to maximize very locally, something that can also be seen when the distance of chosen points to the initially provided input point, $x_{\text{start}}-x_t$ is calculated as shown in Figure~\ref{fig:dist1}.
\FloatBarrier
\noindent
This means that people's behavior in this uni-variate function optimization experiment was based on the attempt of locally maximizing points that they strongly perceived as safe.

\section{Experiment 2: Bivariate functions}
\begin{figure}[ht!]
  \centering
    \includegraphics[scale=0.4]{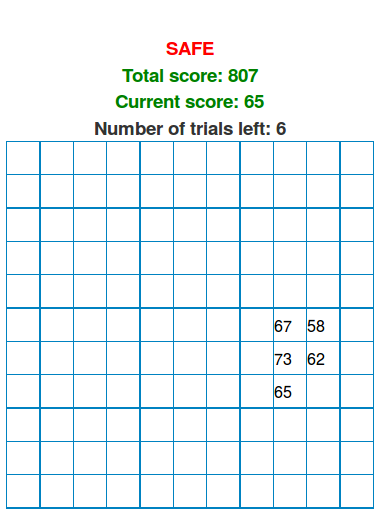}
\caption{Screenshot of second experiment.}
\label{fig:screenshot2}
\end{figure}

In the second experiment, participants were asked to maximize an unknown bivariate function $f: \gvec{x} \rightarrow y$ with $\gvec{x}=(x_1,x_2)^\top$, defined over the grid $x_1, x_2 \in [0,0.05,0.1,\dots,1]$, with $y=f(\gvec{x})+\epsilon$ with $\epsilon \sim \mathcal{N}(0,1)$. As in Experiment 1, the function $f$ was sampled on each block from a $\mathcal{GP}$ with a squared exponential kernel ($l=2$,$\theta=1$). The output values $y$ varied between 0 and 100 and one initial point above 50 was provided. We varied the level of risk within-participants: there were 10 blocks in total out of which 5 were ``normal'', that is unconstrained maximization tasks without a threshold and 5 were ``safe'' blocks in which obtaining an output below 50 caused the current block to end abruptly. The blocks were presented in randomly permuted order. A screenshot is shown in Figure~\ref{fig:screenshot2}.
\subsection{Participants}
62 participants (37 male), with an average age of 31.77 years (SD = 8.97) were recruited via Amazon Mechanical Turk and received \$1 for their participation and a performance-dependent bonus of up to \$1. The average completion time of the whole experiment was 11 minutes.
\subsection{Procedure}
The procedure was essentially the same as for Experiment 1, apart from additional detailed instructions regarding the difference between normal unconstrained (without a threshold) and safe (with a threshold) trials. 
\subsection{Results}
\begin{figure}[ht!]
  \centering
    \includegraphics[scale=0.35]{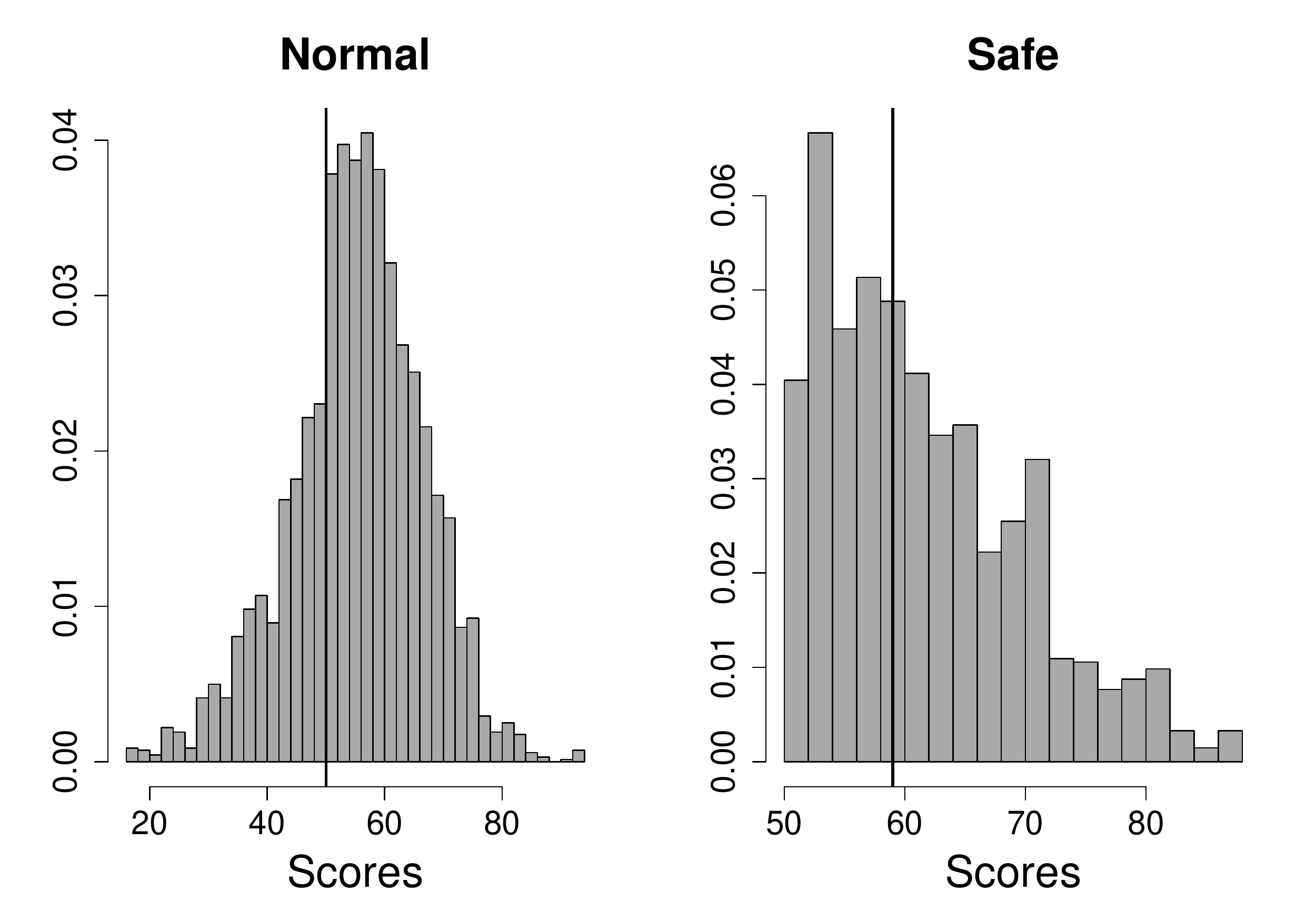}
\caption{Scores per trial.}
\label{fig:raw2}
\end{figure}

As shown in Figure~\ref{fig:raw2}, participants scored better than expected by chance in both the safe and the normal conditions ($t(\text{normal}=50)=24.9$ with $p<0.01$; $t(\text{safe}=59)=9.3$ with $p<0.01$). The reason why chance level performance is higher in the safe condition is that scores below 50 were not allowed and therefore the output was truncated to be above 50.
This time, participants within the safe condition did not complete more trials in a block than expected by randomly choosing inputs on the grid ($t(\text{length}=5)=-0.32$ with $p=0.72$).\\
A similar mixed-effects logistic regression analysis as used for Experiment 1 (Table~\ref{results2}) showed that participants seemed to care most about scoring above the threshold in both conditions. As expected, this effect was more pronounced in the safe conditions than in the normal conditions. Still, the presence of this effect in the normal condition is interesting as it did not matter whether or not participants scored below the threshold. If scoring below 50 did not matter, participants should have not cared as much about sampling above this point in the normal condition as they actually did. One explanation for this might be a transfer effect by which participants assume that sampling below 50 is generally bad. The maximizer set only had a small influence on participants' choices that was slightly bigger for the safe condition. Participants showed no tendency to expand the safe set in either of the conditions. This indicates that most chosen inputs were close to the initial safe input and previously chosen inputs. This relatively high risk aversion is understandable, as the bivariate task is more difficult than the univariate one of Experiment 1.
\begin{table}[ht!]
\centering
\caption{Fixed effects estimates. Significant estimates are flagged.}
\vspace{0.2cm}
\label{results2}
\begin{tabular}{l|lll}
\hline
\textbf{Condition}&\textbf{Variable}& b & SE(b) \\ \hline
\multirow{ 4}{*}{\textbf{Normal}}& Intercept&$-5.17^*$&0.04\\
&Safe Set&$1.35^*$&0.04\\
&Maximizer&$0.13^*$&0.04\\
&Expander&0.04&0.05\\
\hline
\multirow{ 4}{*}{\textbf{Safe}}& Intercept&-5.92$^*$&0.09\\
&Safe Set&$2.11^*$&0.07\\
&Maximizer&$0.23^*$&0.09\\
&Expander&0.03&0.08\\
\end{tabular}
\end{table}
Lastly, a random intercept decision tree analysis (Figure 7) showed that in the best fitting model, only the probability of being above the threshold mattered. This indicates that participants only seemed to care about whether or not an input was safe, simplifying the task to a great extent with a strong focus on the probability of losing. Such simplification makes sense in light of the relative complexity of the bivariate task. 

\begin{figure}[ht!]
\centering
\begin{tikzpicture}
  [
    grow                    = right,
    sibling distance        = 6em,
    level distance          = 9.5em,
    edge from parent/.style = {draw, -latex},
    every node/.style       = {font=\footnotesize},
    sloped
  ]
  \node [root] {root}
    child { node [env] {ignore}
      edge from parent node [below] {$p(sn)<0.8$} }
    child { node [env] {sample}
       edge from parent node [above] {$p(sn)>0.8$} };
\end{tikzpicture}
\caption{Decision tree minimizing log-loss.}
\end{figure}
\begin{figure*}[ht]
  \centering
    \includegraphics[scale=0.3]{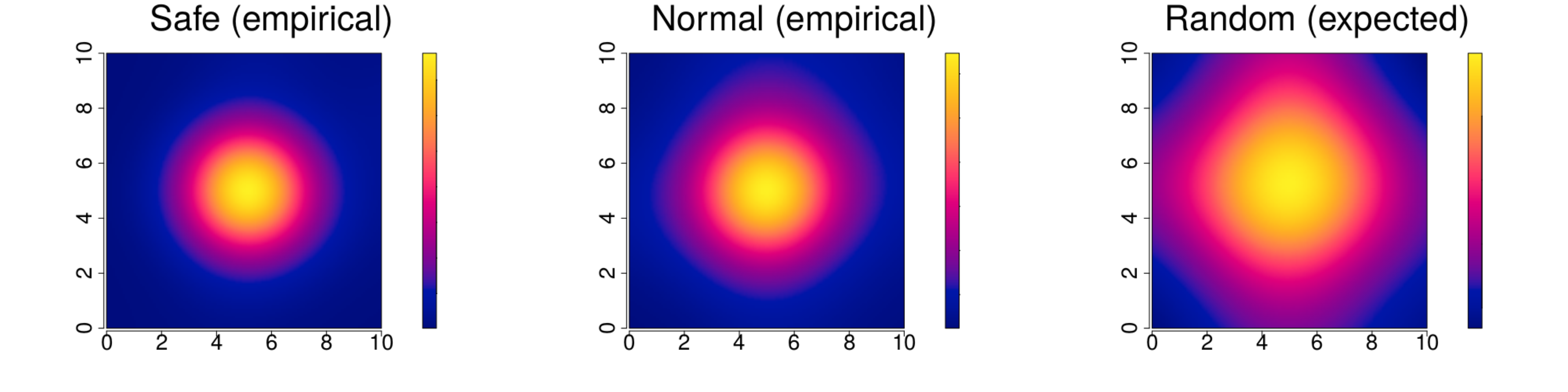}
\caption{Distance of participants' sample points to initial point for safe condition, normal condition, and the theoretically expected distance for random sampling. Participants stay a little closer to the initial point in the safe condition than in the normal condition. Randomly sampling would theoretically cause much higher distances.}
 \label{dist3d}
\end{figure*}

This means that participants again only sampled locally, staying close to the initial point (Figure~\ref{dist3d}).

\FloatBarrier
\noindent
If participants sampled locations are treated as a Poisson process and centred around the initial point, then the posterior density of sampled points shows that participants stayed very close to the initial point in the safe condition, sampled a little further away from the initial point in the normal condition, but never sampled as dispersively as a completely random sampler.

\section{Discussion and Conclusion}

Learning unknown functions and exploiting this knowledge to maximise rewards are essential cognitive skills. Such tasks can be formalized as bandit tasks and here we focused on a restricted version thereof where outcomes below a given threshold need to be avoided. We found that participants' behavior was described well by a Gaussian Process safe optimization routine that establishes safe sets and then tries to maximize outputs within these sets. Participants mostly ignored input points that could expand the safe set, shunning risks and maximizing outputs locally, thereby preferring to rather be ``safe than sorry''.\\
Participants' behavior was consistent with a sequential heuristic in which they first determined whether inputs were safe and then maximized within this safe set. While this strategy involves only local searches, it can result in truly auspicious behavior, especially when the choices are limited. Participants' focus on avoiding unsafe inputs is consistent with a biological homeostasis maintenance principle that prioritizes not loosing everything over gaining as much as possible. The continued influence of the save threshold on participants' choices in the normal condition, where it had no effect on their potential earnings, might be due to participants generalizing their evaluation of outputs below the threshold as ``bad'' from the safe conditions.\\
In future work,  we want to focus on what factors drive participants to switch from explorative to safe behavior and in which situations switching constitutes as a normative strategy, for example because it is minimizing costs \citep{bach2015anxiety}. As we have only focused on functions sampled from a squared exponential kernel here, both for the description of participants' intuitive function learning process and for the actual functions sampled within the task, another direction is to assume different kernel parametrizations of these functions as those lead to diverse theoretical predictions about how fast participants are able to learn \citep{schulzassessing}. Future work could also extend our approach to active versions of more traditional models such as heuristics and weight-based strategies \citep{parpart2015active}.\\
Unlike previous work on human behavior in the bandit setting, which has focused on pure optimization primarily, our work explored a relatively novel facet--optimizing risky functions. We expect that this new approach will provide further insights into how people resourcefully optimize outcomes in the real world.
\bibliographystyle{apa-good}
\bibliography{Bibo}

\begin{thebibliography}{15}
\expandafter\ifx\csname natexlab\endcsname\relax\def\natexlab#1{#1}\fi
\expandafter\ifx\csname url\endcsname\relax
  \def\url#1{{\tt #1}}\fi
\expandafter\ifx\csname urlprefix\endcsname\relax\def\urlprefix{URL }\fi

\bibitem[{Bach(2015)}]{bach2015anxiety}
Bach, D.~R. (2015).
\newblock {A}nxiety-like behavioural inhibition is normative under
  environmental threat-reward correlations.
\newblock {\em PLOS Computational Biology\/}, {\em 11\/}, e1004646.

\bibitem[{Berkenkamp et~al.(2015)Berkenkamp, Schoellig, \&
  Krause}]{berkenkamp2015safe}
Berkenkamp, F., Schoellig, A.~P., \& Krause, A. (2015).
\newblock Safe controller optimization for quadrotors with {G}aussian
  {P}rocesses.
\newblock {\em arXiv preprint arXiv:1509.01066\/}.

\bibitem[{Brochu et~al.(2010)Brochu, Cora, \& De~Freitas}]{brochu2010tutorial}
Brochu, E., Cora, V.~M., \& De~Freitas, N. (2010).
\newblock A tutorial on {B}ayesian optimization of expensive cost functions,
  with application to active user modeling and hierarchical reinforcement
  learning.
\newblock {\em arXiv preprint arXiv:1012.2599\/}.

\bibitem[{Griffiths et~al.(2009)Griffiths, Lucas, Williams, \&
  Kalish}]{griffiths2009modeling}
Griffiths, T.~L., Lucas, C., Williams, J., \& Kalish, M.~L. (2009).
\newblock Modeling human function learning with {G}aussian {P}rocesses.
\newblock In {\em Advances in Neural Information Processing Systems\/}, (pp.
  553--560).

\bibitem[{Huys et~al.(2012)Huys, Eshel, O’Nions, Sheridan, Dayan, \&
  Roiser}]{huys2012bonsai}
Huys, Q., Eshel, N., O’Nions, E., Sheridan, L., Dayan, P., \& Roiser, J.~P.
  (2012).
\newblock Bonsai trees in your head: {H}ow the {P}avlovian system sculpts
  goal-directed choices by pruning decision trees.
\newblock {\em PLoS Computational Biology\/}, {\em 8\/}, e1002410.

\bibitem[{Huys et~al.(2015)Huys, Lally, Faulkner, Eshel, Seifritz, Gershman,
  Dayan, \& Roiser}]{huys2015interplay}
Huys, Q.~J., Lally, N., Faulkner, P., Eshel, N., Seifritz, E., Gershman, S.~J.,
  Dayan, P., \& Roiser, J.~P. (2015).
\newblock Interplay of approximate planning strategies.
\newblock {\em Proceedings of the National Academy of Sciences\/}, {\em 112\/},
  3098--3103.

\bibitem[{Korn \& Bach(2015)}]{korn2015maintaining}
Korn, C.~W., \& Bach, D.~R. (2015).
\newblock {M}aintaining homeostasis by decision-making.
\newblock {\em PLOS Computational Biology\/}, {\em 11(5):e1004301\/}.

\bibitem[{Parpart et~al.(2015)Parpart, Schulz, Speekenbrink, \&
  Love}]{parpart2015active}
Parpart, P., Schulz, E., Speekenbrink, M., \& Love, B.~C. (2015).
\newblock Active learning as a means to distinguish among prominent decision
  strategies.
\newblock In {\em Proceedings of the Thirty-Seventh Annual Conference of the
  Cognitive Science Society\/}, (pp. 1829--1834).

\bibitem[{Rasmussen(2006)}]{rasmussen2006gaussian}
Rasmussen, C.~E. (2006).
\newblock Gaussian {P}rocesses for machine learning.

\bibitem[{Schulz et~al.(2015{\natexlab{a}})Schulz, Konstantinidis, \&
  Speekenbrink}]{schulzexploration}
Schulz, E., Konstantinidis, E., \& Speekenbrink, M. (2015{\natexlab{a}}).
\newblock {E}xploration-exploitation in a contextual multi-armed {B}andit
  {T}ask.
\newblock {\em Proceedings of the 13th International Conference on Cognitive
  Modeling. Groningen, NL\/}.

\bibitem[{Schulz et~al.(2015{\natexlab{b}})Schulz, Konstantinidis, \&
  Speekenbrink}]{schulzlearning}
Schulz, E., Konstantinidis, E., \& Speekenbrink, M. (2015{\natexlab{b}}).
\newblock Learning and decisions in contextual multi-armed bandit tasks.
\newblock {\em Proceedings of the 37th annual conference of the cognitive
  science society\/}, (pp. 2122--2127).

\bibitem[{Schulz et~al.(2015{\natexlab{c}})Schulz, Tenenbaum, Reshef,
  Speekenbrink, \& Gershman}]{schulzassessing}
Schulz, E., Tenenbaum, J.~B., Reshef, D.~N., Speekenbrink, M., \& Gershman,
  S.~J. (2015{\natexlab{c}}).
\newblock {A}ssessing the perceived predictability of functions.
\newblock {\em Proceedings of the 37th annual conference of the cognitive
  science society\/}, (pp. 2116--2121).

\bibitem[{Sela \& Simonoff(2011)}]{sela2011reemtree}
Sela, R., \& Simonoff, J. (2011).
\newblock Reemtree: Regression trees with random effects.
\newblock {\em R package version 0.90\/}, {\em 3\/}, 741--749.

\bibitem[{Srinivas et~al.(2009)Srinivas, Krause, Kakade, \&
  Seeger}]{srinivas2009gaussian}
Srinivas, N., Krause, A., Kakade, S.~M., \& Seeger, M. (2009).
\newblock Gaussian {P}rocess optimization in the bandit setting: {N}o regret
  and experimental design.
\newblock {\em arXiv preprint arXiv:0912.3995\/}.

\bibitem[{Sui et~al.(2015)Sui, Gotovos, Burdick, \& Krause}]{sui15icml}
Sui, Y., Gotovos, A., Burdick, J.~W., \& Krause, A. (2015).
\newblock Safe exploration for optimization with {G}aussian {P}rocesses.
\newblock In {\em International Conference on Machine Learning (ICML)\/}.

\end{thebibliography}
\end{document}